\documentclass{article}
\usepackage[utf8]{inputenc}
\usepackage{authblk}
\usepackage{setspace}
\usepackage[margin=1.25in]{geometry}
\usepackage{graphicx}
\usepackage{subcaption}
\usepackage{amsmath}
\usepackage{amssymb}
\usepackage{booktabs}
\usepackage{lineno}
\usepackage{multirow}
\usepackage{colortbl}
\usepackage{xcolor}
\usepackage{makecell}
\usepackage{threeparttable}
\usepackage{comment}
\usepackage{siunitx}
\usepackage{hyperref}
\newcommand{\fref}[1]{Fig.\ {\ref{#1}}}
\newcommand{\tref}[1]{Table\ {\ref{#1}}}

\usepackage[style=nature]{biblatex}
\title{Magnetic field estimation using Gaussian process regression for interactive wireless power system design}

\author[1]{*Yuichi Honjo}
\author[1]{Cedric Caremel}
\author[1]{Ken Takaki}
\author[1,2]{Yuta Noma}
\author[1]{Yoshihiro Kawahara}
\author[1]{*Takuya Sasatani}

\affil[1]{Department of Electrical Engineering and Information Systems, The University of Tokyo, 7-3-1 Hongo, Bunkyo-ku, Tokyo 113-8656, Japan}
\affil[2]{Department of Computer Science, University of Toronto, 40 St George St, Toronto, ON, M5S 2E4, Canada}
\affil[ ]{Email: honjo@akg.t.u-tokyo.ac.jp, sasatani@g.ecc.u-tokyo.ac.jp}

\date{}

\begin{document}
\maketitle

%%%%%% Abstract %%%%%%
\begin{abstract}
Wireless power transfer (WPT) with coupled resonators offers a promising solution for the seamless powering of electronic devices. Interactive design approaches that visualize the magnetic field and power transfer efficiency based on system geometry adjustments can facilitate the understanding and exploration of the behavior of these systems for dynamic applications. However, typical electromagnetic field simulation methods, such as the Method of Moments (MoM), require significant computational resources, limiting the rate at which computation can be performed for acceptable interactivity. Furthermore, the system's sensitivity to positional and geometrical changes necessitates a large number of simulations, and structures such as ferromagnetic shields further complicate these simulations. Here, we introduce a machine learning approach using Gaussian Process Regression (GPR), demonstrating for the first time the rapid estimation of the entire magnetic field and power transfer efficiency for near-field coupled systems. To achieve quick and accurate estimation, we develop 3D adaptive grid systems and an active learning strategy to effectively capture the nonlinear interactions between complex system geometries and magnetic fields. By training a regression model, our approach achieves magnetic field computation with sub-second latency and with an average error of less than 6\% when validated against independent electromagnetic simulation results.
\end{abstract}

%%%%%% Main Text %%%%%%

\section*{Introduction}
Wireless power transfer (WPT) using magnetically coupled resonators offers a promising solution for powering electronic devices. This technology is particularly useful in dynamic applications, such as electric vehicles and mobile electronics\cite{WPT-magnetic-resonance-1, WPT-overview, WPT-body-coupled, WPT-roomscale, WPT-medical, WPT-EV-WPT-system-2}. In these areas, evaluating system performance and electromagnetic field distribution is critical to ensure safety and minimize interference with nearby electronic devices across various possible positions. Developing interactive design approaches that compute and visualize magnetic field exposure and power transfer efficiency based on system geometry adjustments, such as component geometry and placement, would be highly beneficial for providing insights into system behavior. However, analyzing near-field coupling scenarios presents substantial computational challenges. These challenges arise from the complexity and intensive computation needed to accurately characterize interactions between multiple couplers, often relying on resource-demanding methods such as the Finite Element Method (FEM) and the Method of Moments (MoM)~\cite{WPT-exposure-system-design-1, WPT-human-exposure, WPT-magnetic-resonance-2, WPT-coil-design}. These methods, while comprehensive, are unsuitable for real-time, iterative computations. Recent analytical approaches reduce computational demands by constraining the solution space to geometries that admit simplified or empirically derived solutions, but this limits their applicability to a narrow set of predefined structures~\cite{AnalyticalApproaches1, AnalyticalApproaches2, AnalyticalApproaches3}.

The computational demands associated with physics simulations have driven the exploration of machine learning and regression techniques for rapid estimation~\cite{Learning-magnetic-field-estimation, Learning-fluid-1, Learning-fluid-umetani-2, Learning-EV-WPT-system, Learning-possion, Learning-WPT-gaussian, Learning-audio2, Learning-nerf2, Learning-carbon-nanotube, Learning-climate, Learning-fluid-3, Learning-conformal-system}. In the field of electromagnetics, these techniques have successfully estimated low-dimensional metrics in near-field systems, such as power transfer efficiency in WPT~\cite{Learning-EV-WPT-system, Learning-WPT-gaussian}, and can provide field estimations for far-field systems by leveraging the decoupled nature of transmitters and receivers~\cite{Learning-audio2, Learning-nerf2}. However, estimation of the entire near-field vector representation using machine learning remains underexplored. While computational design approaches for ferromagnetic shields have been proposed, accommodating numerous potential device placements poses a significant challenge~\cite{Learning-optimization-WPT}. Achieving precise estimations of comprehensive physical fields in near-field systems, where multiple dynamic components significantly influence the field, is especially difficult, and the inclusion of ferromagnetic shields adds further complexity. In parallel, recent advances in fluid dynamics research demonstrated the potential of using Gaussian Process Regression (GPR) for rapidly and accurately computing complex interactions between physical objects and volumetric fields~\cite{Learning-fluid-umetani-2, GPR-1, GPR-2, GPR-3}.

Here, we show that WPT systems based on near-field coupling can also be modeled using the GPR method due to the similarity in the highly nonlinear interactions between the coupled resonators' geometry and their magnetic field. To the best of our knowledge, this work is the first to present a machine-learning strategy for interactively estimating entire magnetic fields. We propose a GPR pipeline specifically tailored for our WPT systems to rapidly and accurately estimate magnetic field distribution and power transfer efficiency based on shield geometry and transmitter-receiver positions. Straightforward application of machine learning techniques can lead to unstable and inaccurate estimations due to highly nonlinear input-output relationships and typically requires extensive simulation training datasets. To address these challenges, we enhance the training data points collection process and improve estimation accuracy by customizing the modeling, computation, and training data points acquisition processes to fit the unique characteristics of dynamic WPT system design. This enables sub-second estimation of magnetic fields and power transfer efficiency, facilitating interactive and intuitive WPT system design.

The main challenges in developing effective estimation models for magnetically-coupled WPT systems include (i) the highly nonlinear variability of the magnetic field with changes in shield geometry and position and (ii) the resource-intensive training data points collection required for each transmitter-receiver configuration. To address the first challenge, we develop a geometry-based adaptive exterior grid system inspired by computer graphics techniques for modeling stretching skin \cite{LBS-linear-blend-skinning}, which, in our case, stretches the field model to alleviate rapid changes and helps linearize the system input-output relationships. Additionally, we propose an aligned-edge polycube parameterization, adapted from the original polycube method known for expressing smooth geometries in fluid dynamics \cite{Learning-fluid-umetani-2}. Our approach is tailored to modeling geometries with angular edges and for systematically and iteratively deforming arrays of voxels to represent shield geometry in a regression-compatible, fixed-dimension format. To address the second challenge and efficiently minimize the required amount of training data points, we implement active learning based on GPR variance, to identify the critical training data points that will enhance the regression model. Furthermore, we explore data preprocessing and postprocessing strategies to improve the problem's linearity and estimation accuracy.

Altogether, these innovations enable rapid and precise estimations of magnetic field distribution, transferable power, and power transfer efficiency. Using our trained GPR model, our evaluations demonstrate field estimation within \SI{130}{ms}, a 13,523-fold speed improvement compared to MoM electromagnetic simulations (Supplementary video 1). This performance is validated with less than a 6\% average error against independent electromagnetic simulation results. As detailed in Supplementary Table~1, our method allows for the interactive estimation of magnetic fields across various system geometries with significantly fewer training data points, a task that posed challenges with conventional machine learning approaches. This advancement empowers users to gain profound insights into the complex behavior of dynamic WPT systems and facilitates intuitive exploration of system behavior that conventional electromagnetic solvers cannot easily achieve.

\section*{Results}
\begin{figure}[h]
    \centering
    \includegraphics[width=176.5mm]{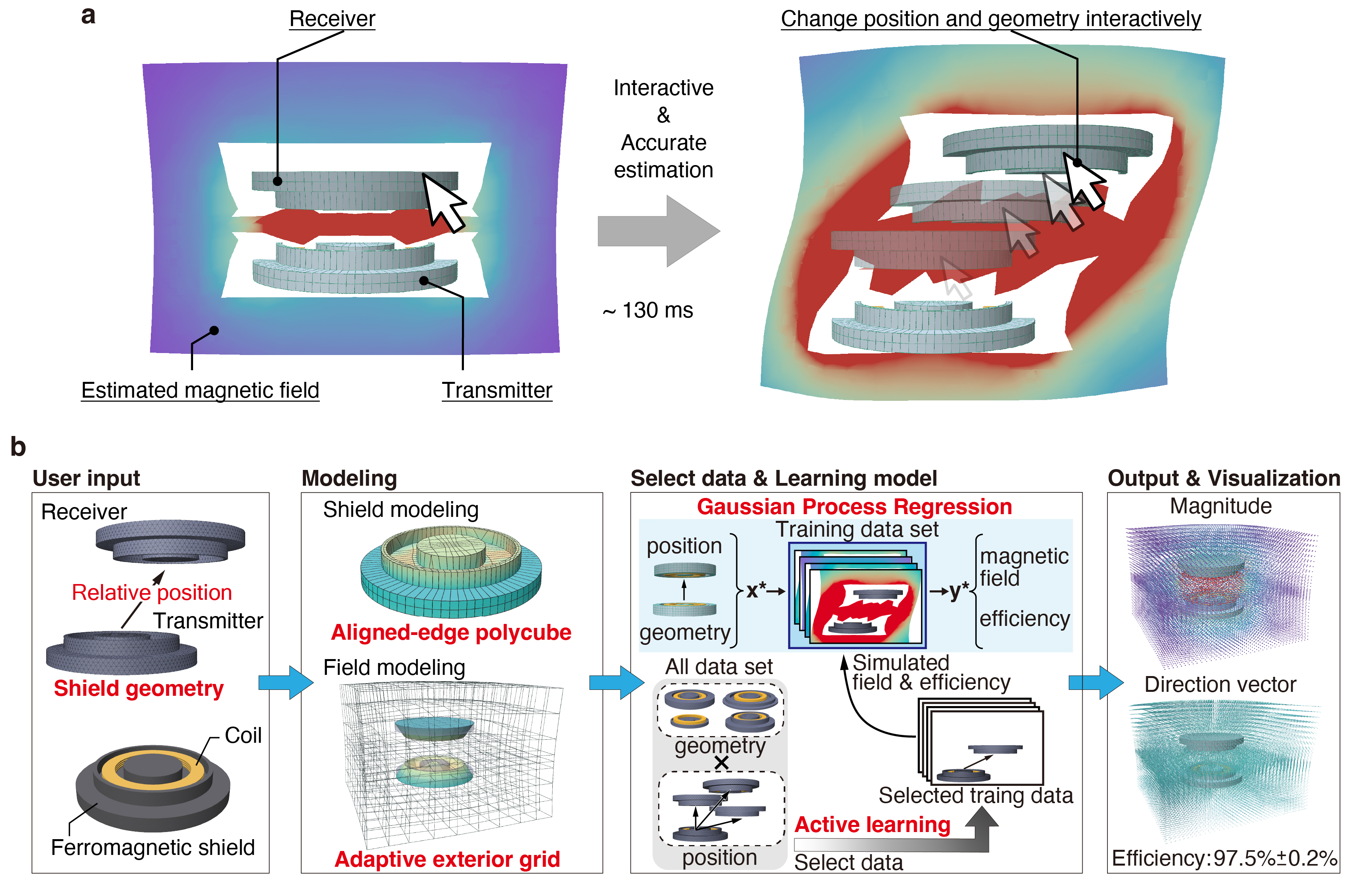}
    \caption{Magnetic field estimation in a wireless power transfer system and its estimation pipeline.
    \textbf{a} The magnetic field in a wireless power transfer system is rapidly estimated using machine learning techniques. This approach facilitates interactive exploration and design of system geometries by adjusting the shield's position and geometry.
    \textbf{b} An overview of the estimation method. User input includes details about the shield geometry and the relative positioning of the transmitter and receiver. For wireless power transfer, we use coils equipped with a ferromagnetic shield. The shield geometries are represented using the proposed aligned-edge polycube mesh, which accommodates various geometries expressed in fixed-length compact vectors. Furthermore, estimation points are modeled using our proposed adaptive exterior grid. This grid alleviates nonlinearities and discontinuous relationships between geometric inputs and magnetic field outputs, essential for precise estimation via GPR. The GPR model's training data is efficiently curated using active learning, enabling the development of reliable models with fewer data points. This method allows for the rapid estimation and visualization of the magnetic field's magnitude and vector, as well as power transfer efficiency.}
    \label{fig:overview}
\end{figure}

This study aims to estimate the magnetic field and power transfer efficiency of a dynamic WPT system, wherein the receiver can be positioned anywhere within a specified range, as shown in \fref{fig:overview}a. This system transfers power from the transmitter to the receiver via an oscillating magnetic field generated by a current applied to the transmitter coil shown in Supplementary Fig.~1a. In practical applications, ferromagnetic shields are often used to shape and suppress magnetic field leakage. These shields are essential for maintaining power distribution within exposure limits and defining keep-out zones for high-power applications, such as electric vehicle (EV) charging~\cite{WPT-EV-WPT-system-2}. Our estimation approach facilitates shield design and allows for an interactive examination of how external magnetic fields change with different receiver placements. Further details of this setup are described in the Methods section.

The core of our approach to learning-based magnetic field estimation comprises two main stages: (i) a preprocessing, parameterization stage, where the input-output relationship is linearized and (ii) a model training stage, along with some postprocessing, regressing training dataset generated from FEKO, a commercial Method of Moments-based electromagnetic field solver (see methods). We later develop a visualization stage with a GUI interface for interacting with the trained model, as illustrated in \fref{fig:overview}b. Regarding the first parameterization stage, prior work demonstrated that improving the problem's linearity greatly enhances the estimation accuracy~\cite{Learning-fluid-umetani-2}. Regarding the second stage, when selecting our learning approach, we prioritized minimizing the amount of physical simulation data required for model training, since each simulation is both time- and resource-intensive. This consideration led us to choose the Gaussian Process Regression method, a technique capable of representing complex relationships efficiently by leveraging the variance within the training data points to identify critical training data points that can contribute to the final estimation with maximum accuracy. Importantly, this method has been deployed successfully for real-time interactions in fluid dynamics, and has been confirmed to be more accurate than other machine learning methods when only a small amount of training data point is available~\cite{Learning-fluid-umetani-2}.

The following sections detail the methodology we have developed, including geometry-based adaptive grid systems and our model training approach, which leverages active learning with GPR variance. We then perform the performance evaluations and discuss our work contributions. The methods section provides details on the training dataset input, GPR model and visualization stage.

\subsection*{Preprocessing: 3D geometry parameterization}
Our strategy for geometry parameterization involves modeling shield structures to accurately represent various shield geometries while adjusting points on a 3D grid so that the magnetic field can be correctly sampled. This adjustment is intended to improve the continuity and linearity of the estimation process, which is crucial because ferromagnetic shields can cause discontinuities in the magnetic field; for example, if a grid point moves in and out of the shielded volume, the sampled field value can be significantly affected.
We first describe our approach for modeling shield geometries and, based on this, detail our method for adapting exterior grid points to ensure continuous changes in the sampled field.

\subsubsection*{Expressing complex shield geometries by an aligned-edge polycube approach}
Incorporating three-dimensional geometries into machine learning for efficient regression presents significant challenges and often requires customized modeling techniques. Traditional 3D geometry modeling approaches, such as integrating multiple views \cite{multiple-view}, voxel-based modeling \cite{voxel-based-modeling-1, voxel-based-modeling-2}, and point cloud data of surfaces \cite{Modeling-pointnet}, produce discrete responses to continuous geometry changes or different dimensions for each geometry, making them unsuitable for regression tasks. The polycube parameterization was initially developed for estimating fluid dynamics by iteratively adapting arrays of voxels to fit various smooth geometries, a better-suited process for regression \cite{Learning-fluid-umetani-2}. However, its direct application would be inadequate for representing angular geometries (often used in the construction of shields in WPT systems). To address this, we develop a new method, the aligned-edge polycube approach (Supplementary video~2).

\begin{figure}[h]
    \centering
    \includegraphics[width=176.5mm]{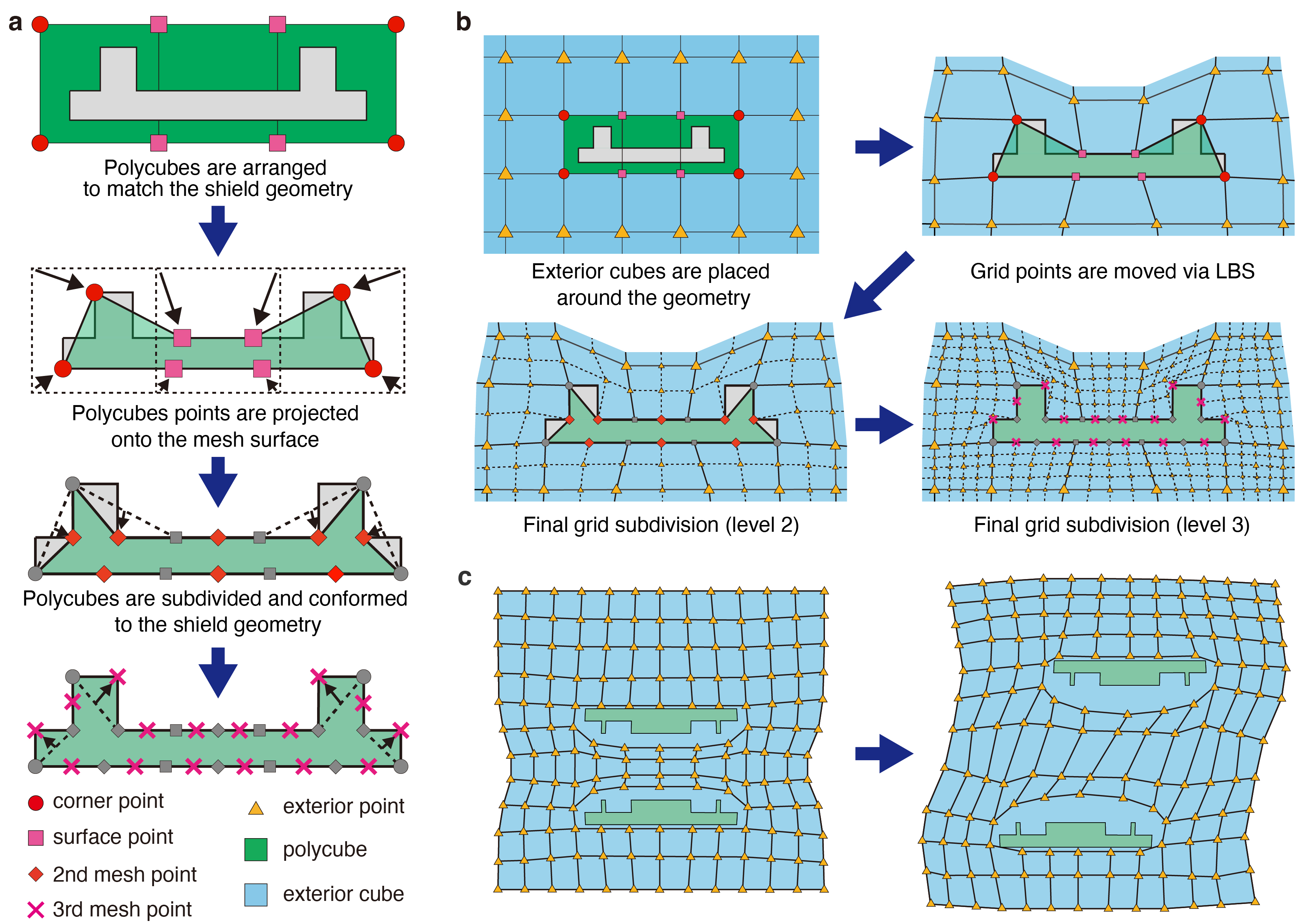}
    \caption{Aligned-edge polycube and adaptive exterior grid.
    \textbf{a} Cross-sectional view for generating the aligned-edge polycube. The polycubes (green) are arranged around the shield geometry (gray) and are fitted to the surface of the geometry. Initial points consist of corner points and surface points forming shield-fitting cubes. The mesh is subdivided by interpolating points. By aligning them to the edges of the geometry, the aligned-edge polycube minimizes the number of points needed to represent the shield geometry.
    \textbf{b} Generation process for the adaptive exterior grid. Exterior cubes (blue) are positioned around shield-fitting polycubes and adjusted to match the geometry representation. These polycubes are further subdivided by interpolation relative to the exterior cubes.
    \textbf{c} Adjusting the adaptive exterior grid to coil positioning. It is modified to maintain its relative position to the coils, taking into account the transmitter and receiver's relative positions. This adjustment helps prevent abrupt changes in the magnetic field due to position variations.}
    \label{fig:polycubemesh}
\end{figure}

The aligned-edge polycube mesh effectively represents complex angular geometries by arranging multiple cubes around the 3D model of a target shield, projecting their vertices onto the shield geometry, and fitting the vertices to the geometry corners (\fref{fig:polycubemesh}a). The green cubes in \fref{fig:polycubemesh}b represent the aligned-edge polycube expression of the shield geometry, whereas the blue ones represent the adaptive exterior grid described in the next section. This allows for a fixed-vector input (a prerequisite for our model) and the accurate representation of shield structures with minimal parameters by iteratively segmenting surface voxels and aligning them to their corners (Supplementary Fig.~2). This projection offers several advantages that enhance training simplicity and estimation accuracy: (i) subdivision of cubes, categorized from level 1 to 3, in order to increase the point count and ensure vector length consistency regardless of geometry variations; (ii) alterations of a part of the geometry do not affect meshes in other regions, allowing consistent meshing for unchanged regions. Other advantages compared with other methods are shown in Supplementary Table~2.

\subsubsection*{Enhancing input-output linearity by adaptive exterior grid sampling}
Magnetic field estimations on a fixed grid often suffer inaccuracies, particularly with position changes and geometric modifications of the target coil within a shield. Minimal changes can cause drastic shifts in weak magnetic field regions, especially if obstructed by an opposing ferrite shield, leading to discontinuous changes in the magnetic field that will further complicate modeling. To address this, our method instead adapts the sampling grid relative to the coil's position (Supplementary video~3).

\begin{figure}[h]
    \centering
    \includegraphics[width=84mm]{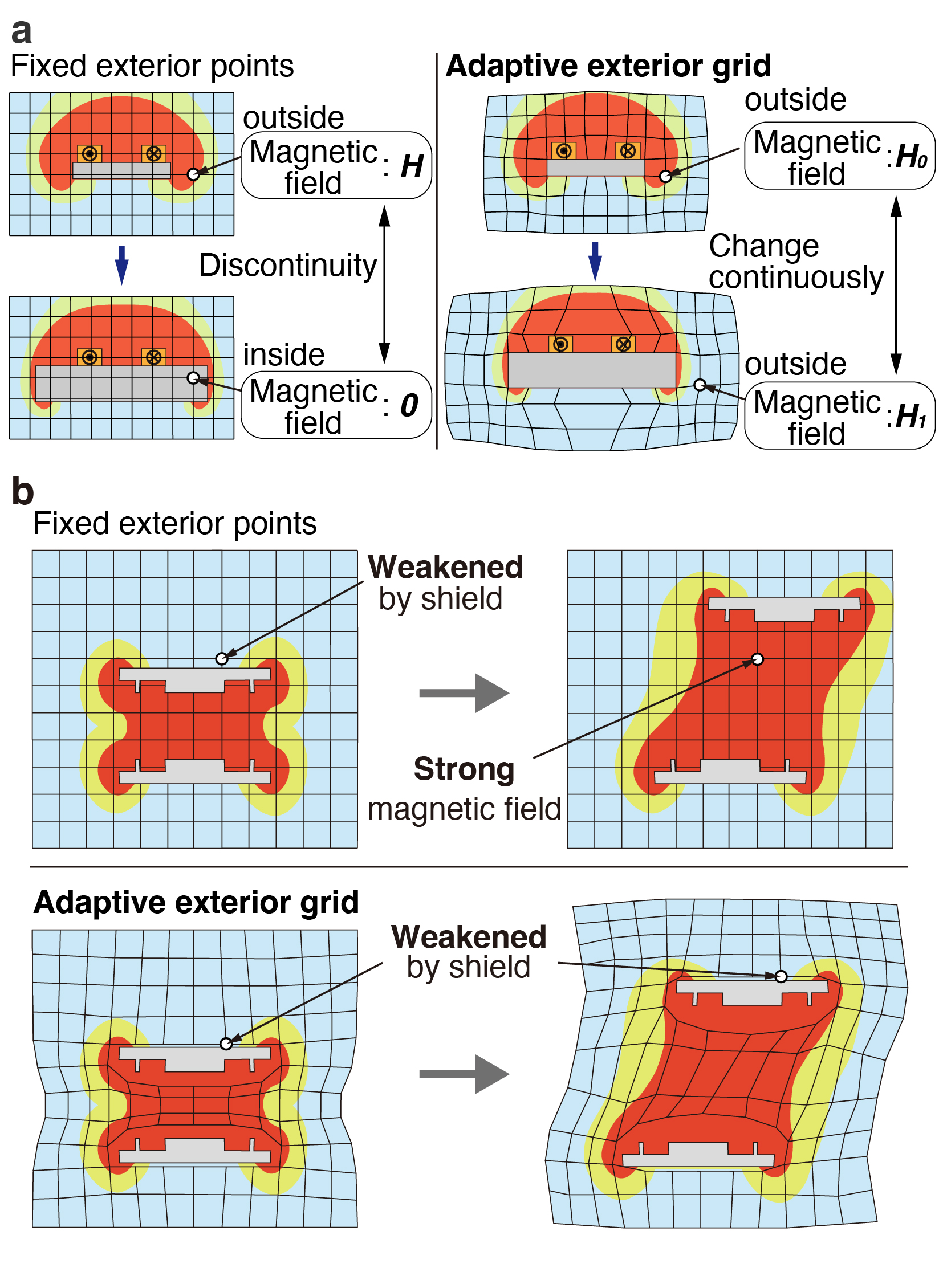}
    \caption{Principle of continuity and linearity enhancement using an adaptive exterior grid.
    \textbf{a} Shows the fixed grid estimation point transitioning between inside and outside the ferromagnetic shield, leading to discontinuous changes in the magnetic field. The proposed adaptive grid, on the other hand, adjusts its position in response to changes in geometry, thereby maintaining continuity.
    \textbf{b} Demonstrates how changes in shield position can cause discontinuous and non-linear variations in the magnetic field when using fixed exterior grids. The proposed adaptive exterior grid mitigates these effects by adapting to the geometry, thereby minimizing changes in the magnetic field magnitude at the estimated points.}
    \label{fig:linearity}
\end{figure}

Our adaptive exterior grid strategy, akin to spatial modeling, places voxels outside the resonator geometry. These grid points are adjusted to align with the coil geometry and subdivided as much as necessary (\fref{fig:polycubemesh}b). The grid adapts to the positions of the transmitter and receiver, as shown in \fref{fig:polycubemesh}c, ensuring improvements in the continuity and linearity of input-output relationships as the shield geometry changes (\fref{fig:linearity}a). This approach allows for flexible estimation points in response to shield position modifications as well (\fref{fig:linearity}b). Detailed advantages over other grid approaches are provided in  Supplementary Table~3.

\fref{fig:polycubeResult} illustrates the application to shield geometry of these aligned-edge polycube mesh and adaptive exterior grid strategies. The adaptive grid is sufficiently detailed from subdivision level 2, while subdivision level 3 is required for the polycube mesh to effectively capture the targeted geometry.

\subsection*{Model training}
We start with an initial training dataset consisting of 744 shields, simulated using FEKO, with positions interpolated on a common grid to ensure that the solution space is consistent over all training data points (step 1 and 2 in \fref{fig:activelearning}a). Using this initial training data set, we select six exterior points from the exterior grid level 1 (the details are explained in the Method section). Instances of our GPR model at each of the six pre-selected exterior points are then trained based on this initial training set (step 3 in \fref{fig:activelearning}a). In the active learning process, from the unlabeled data, our method determines points of interest, based on the highest variance at each of those six exterior points, so that the model's accuracy can be enhanced (step 4 and 5 in \fref{fig:activelearning}a). Additionally,  thanks to our FEKO simulations, we qualify this high-variance data as `labeled data', and re-train instances of our GPR model (step 6 in \fref{fig:activelearning}a). We repeat this process until the initial dataset is doubled (a dataset of now 1,488 labeled data), increasing the number of points we can confidently rely on for our model training. Finally, in a postprocessing phase, we use the power transformation to avoid abrupt changes and linearize the relationship between input and output, further improving estimation accuracy.

\subsubsection*{Gaussian process regression model}

The input data $\mathbf{x}$ for GPR consists of the coordinates of points on the aligned-edge polycube mesh, \textit{i.e.}, 1122 positions $P_i(x,y,z)$, and the relative positions of the transmitter and receiver $P_{\rm TX-RX}(y,z)$, totaling 3368 values. The output $y^*$ corresponds to the estimated magnetic field at each point, along with an estimation of the power transfer efficiency.

In GPR, the kernel matrix is defined using inputs $\mathbf{x}$, $\mathbf{x}^\prime$ and a kernel function $k_\mathrm{GPR}(\mathbf{x}, \mathbf{x}^\prime)$.
\begin{linenomath}
\begin{align}
    \mathbf{K_{GPR}}_{nn^{\prime}} =k_\mathrm{GPR}(\mathbf{x}^n,\mathbf{x}^{n^{\prime}})
\end{align}
\end{linenomath}
Details about the kernel function are elaborated in the Method Section. The probability distribution of estimations for new inputs follows a Gaussian distribution:
\begin{linenomath}
\begin{align}
    p(y^*|\mathbf{x}^*,\mathcal{D}) &= \mathcal{N}(\mathbf{k_{GPR}}_*^T\mathbf{K_{GPR}}^{-1}\mathbf{y}, k_\mathrm{GPR**}-\mathbf{k_{GPR}}_*^T\mathbf{K_{GPR}}^{-1}\mathbf{k_{GPR}}_*) \\
    \mathbf{k_{GPR}}_* &= (k_\mathrm{GPR}(\mathbf{x}^*,\mathbf{x}^1), k_\mathrm{GPR}(\mathbf{x}^*,\mathbf{x}^2), \cdots, k_\mathrm{GPR}(\mathbf{x}^*,\mathbf{x}^N))^T \\
    k_\mathrm{GPR**} &= k_\mathrm{GPR}(\mathbf{x}^*,\mathbf{x}^*)
\end{align}
\end{linenomath}
Here, $\mathcal{D}$, $\mathbf{x}^*$, $\mathbf{y}^*$ represent the trained model, new input, and estimated output, respectively (note that the final estimation vector \(\mathbf{y}^*\) is a concatenation of elements \(y^*\), each corresponding to either a prediction point or the power transfer efficiency scalar).

\subsubsection*{Active learning using GPR variance}
Active learning is an efficient technique for creating an initial training dataset, when only a small amount of training data points is available~\cite{Activelearning-bayesian}. In GPR, the choice of training data points is critical, as the estimation accuracy heavily depends on that training data. The training data points are prepared by setting the geometry and relative positioning of the transmitter-receiver couple, conducting electromagnetic field simulations, and generating accurate data. Since these simulations are computationally expensive, preparing a large volume of training data points is challenging, necessitating the efficient selection of essential data points. Active learning therefore assists in determining which data points to incorporate.

\fref{fig:activelearning}a provides an overview of the active learning process used in this study. We generated data for 169 relative positions for each of the 372 shields, resulting in 62,868 unlabeled data points. The initial training dataset was collected by executing electromagnetic simulations at interpolated grid positions so the collected data would exist throughout the solution's space. We extracted 744 initial training dataset from this, with the remaining points forming our unlabeled dataset. Electromagnetic simulations are conducted using FEKO for the labeled training dataset, which is then used to train the GPR model for six estimation points. The parameterized unlabeled dataset is input into the trained GPR model, and deviations for each data point are calculated.

These deviations, derived from a Gaussian distribution, indicate the probability range (approximately 68\% or 1 $\sigma$) where the solution lies; larger deviations imply greater estimation difficulty. For the training set, we also select unlabeled data points that corresponds to a large deviation for each estimation point. As this process involves six estimation points, six unlabeled data are added per cycle. This cycle is repeated until the training set doubles, reaching 1,488 data points.

\subsubsection*{Postprocessing linearization}
To further increase accuracy, we apply a transformation to the training data values used in GPR as follows:

\begin{linenomath}
\begin{align}
        y^* = \Bigl(\mathbf{k_{GPR}}_*^T\mathbf{K_{GPR}}^{-1}\mathbf{y}^{\frac{1}{s}}\Bigr)^s
\end{align}
\end{linenomath}
This transformation helps make skewed data more closely resemble a normal distribution~\cite{Transformation-variable}. Complex transformations such as the Box-Cox transformation~\cite{Transformation-boxcox} and the Yeo-Johnson transformation~\cite{Transformation-yeojohnson} are normally used, but the inverse transformation would be non-trivial~\cite{Transformation-variable}. In this study, we use the power transformation, as seen in equation (5), such that its inverse transformation is easily tractable. In this transformation, we raise the value of $\mathbf{y}$ to the power of $1/s$ and then restore the calculated estimation by raising it to the power of $s$.
By doing this transformation, it is possible to improve the estimation accuracy where the change would be too large to be represented by the GPR model, as shown in \fref{fig:pow}b.
A simple example of how this postprocessing step helps improve accuracy is shown in Supplementary Fig.~3.

\subsection*{Evaluation}

\subsubsection*{Preprocessing linearity enhancement by adaptive exterior grid}
\fref{fig:evaluationExteriorGrid} shows the improvement in estimation due to our proposed adaptive exterior grid. This is the result of a cross-validation process, where the 1,488 data points are divided into 12 groups, 11 of which are used for training, and the magnitude and vector of the magnetic field are estimated for the remaining set. This process is repeated 12 times. \fref{fig:evaluationExteriorGrid} shows how the average error for each estimated point improved when the estimation point layout was changed from a fixed grid to an adaptive exterior grid. The mean square error is used for the magnitude, and the absolute error is used for the vector. 

Using the adaptive exterior grid significantly reduces the error for the magnetic field magnitude. The accuracy of the vector estimation is also improved, albeit not to the same extent as the magnitude. This suggests that the adaptive exterior grid maintains the relative positions of the sampled point with reference to the shield, thereby improving the linearity between input and output and making the GPR estimations more accurate.

\subsubsection*{Training dataset quality enhancement by active learning}
\fref{fig:activelearning}b shows the relative positions of the two shields in the labeled initial training dataset before active learning. This dataset is selected to ensure that various shield geometries are placed in various positions. \fref{fig:activelearning}c shows the relative positions of the data points in the training set after active learning. Points of interest are determined by our algorithm in areas where there were previously fewer points. Furthermore, our results show that the selected points are concentrated near the boundaries of the solution space. This can be explained by the fact that GPR calculates the distance from the training data. Since the points near the boundary have fewer nearby data points, they show higher variance, and thus, more points can be selected.

\fref{fig:activelearning}d, e show the maximum deviation and maximum error at each estimated point when the amount of data points is increased by active learning (increase of the amount of training data). The deviation decreases uniformly at all points. Moreover, the error reduction is more significant at points estimated with larger initial errors and decreases at all points as the amount of training data increases. This indicates that active learning efficiently increased the amount of data points from the difficult-to-estimate areas, thereby reducing the estimation error.

\subsubsection*{Estimation accuracy improvement by postprocessing}
Firstly, the magnetic field estimation accuracy according to the value of $s$ is shown in \fref{fig:pow}a and is plotted as the relative error of magnetic field magnitude, the mean of the relative error at each estimation point for 50 test data. 
It also shows the overall average and standard deviation of all plots. These results show that the larger the value of $s$, the better the estimation accuracy and the standard deviation. 
When $s$ is sufficiently large, the error converges to a small value.
The estimation accuracy improves as $s$ increases because the magnetic field changes significantly at some points, as shown in \fref{fig:pow}b.
\fref{fig:pow}b illustrates the value of the magnetic field at one estimation point when the same shield is moved and this can be expressed as $\mathbf{y}^{\frac{1}{s}}$. When $s=1$, the linearity is low in areas with large changes, and the estimation accuracy decreases (Supplementary Fig.~4a). In contrast, using $s=20$ reduces the changes in magnetic field magnitude as shown in \fref{fig:pow}b, consequently improving the linearity and the estimation accuracy. \fref{fig:pow}c shows the same evaluation result for power transfer efficiency. This shows that the average error and standard deviation of the power transfer efficiency estimation are minimized at $s = 1/6$, which differs from the magnitude results. \fref{fig:pow}d shows the comparison of power transfer efficiency for $s=1$ and $s = 1/6$. The overall changes are less steep at $s=1/6$, contributing to superior linearity and estimation accuracy.

\fref{fig:pow}e shows the error map of magnetic field distribution and power transfer efficiency estimations using the value of $s$ that yielded the best results, associated with the field visualizations of one of the harder-to-estimate performance samples. The magnetic field estimation has a relative error of less than 10\% in most places, and even at its maximum, is about 15\%, with an overall average of less than 6\%. As for power transfer efficiency, the relative error is less than 1\% in most places, and even at its maximum, is about 3\%, with an overall average of less than 0.7\%, achieving high estimation accuracy. 
Furthermore, \fref{fig:pow}f, g respectively show the relative error of the estimated magnetic field magnitude and power transfer efficiency using the cumulative distribution function (CFD). For the magnetic field estimation, we show that 80\% of estimations are within 12\% error. Additionally, 80\% of the estimations are within 0.7\% error for the power transfer efficiency. The improvements in relative error accuracy achieved by the proposed methods are summarized in Supplementary Table~4.

\subsubsection*{Computational speed}
In the final stage, we visualize the output results to allow for interactive shield design, the main motivation for this work. The processing time required to estimate the magnetic field distribution in response to changes in the geometry and relative positions of the shields is benchmarked on a MacBook Pro (2023 model, M2 Pro chip) and measured at a maximum of approximately \SI{130}{ms}, enough to achieve sufficiently smooth interactivity.

\tref{table:speed} shows the response times for estimating the magnetic field distribution according to the subdivision level of the exterior grid. The processing time is calculated separately for changes in position and geometry. As the number of estimation points increases proportionally with the subdivision level, the time required for estimation also increases. Notably, at level 3, the time increases significantly (\SI{1287}{ms}) but level 2 (\SI{130}{ms}) is sufficiently fast and accurate for interactive design. 
In comparison, the magnetic field distribution calculation using the FEKO electromagnetic simulator takes an average of \SI{1758}{s} on an Intel(R) Xeon(R) CPU E5-2698 v3 (\SI{2.30}{GHz}, 16 cores, 16 threads), indicating that the proposed method operates a 13,523-fold speed increase compared to traditional MoM simulations.

\section*{Discussion}
\subsection*{Usage scenarios}
To demonstrate the utility of our method, we provide an interactive demo, publicly available on GitHub\footnote{\url{https://github.com/SasataniLab/EMFieldGPR}}\footnote{\url{https://doi.org/10.5281/zenodo.17378263}}. Because running live software is difficult to convey on paper, we present the benefits through case studies that adjust several parameters and report the resulting outputs in Fig.~\ref{fig:complexity}. These examples illustrate what can be adjusted interactively and intuitively while observing outcomes in real time.

In the static wireless power case (Fig.~\ref{fig:complexity}c, d, and e), users can adjust shield dimensions (\textit{e.g.}, $h_2$, $h_3$) while monitoring the magnetic-field distribution and transfer efficiency. This enables interactive exploration of subtle design changes, revealing geometry–performance trade-offs that would be prohibitively slow to evaluate with conventional full-wave simulations.

In the dynamic powering case (Fig.~\ref{fig:complexity}f, g, h, and i), users can track how magnetic fields evolve at points of interest under coil misalignment. Point~A is a fixed reference location on the ground, while Point~B emulates a receiver-side point that moves with the power-receiving device. By observing both efficiency and field strength at these points, designers can identify acceptable misalignment ranges and tune shield parameters (\textit{e.g.}, $r_1$) to suppress unwanted field levels.

While these case studies illustrate only a minimal subset of parameters for clarity, the approach generalizes to a wider range of tunable variables, enabling real-time exploration of complex geometry–performance relationships—capabilities that are impractical with conventional electromagnetic solvers.

\subsection*{Comparison with prior work}
This work demonstrates that our GPR pipeline, tailored for near-field WPT systems, enables rapid and accurate magnetic field estimation, facilitating interactive exploration of dynamic system behavior and intuitive magnetic shield design. Prior machine learning-based field estimation approaches struggled to balance quick computation with accurate analysis due to the significant sensitivity of magnetic fields to subtle changes in system geometry and component positioning. Our proposed grid generation methods and output data shaping enhance the continuity and linearity of the field-geometry relationship. Moreover, the active learning based on GPR variance ensures accurate estimation with minimal dataset. Together, these innovations achieve precise estimation at interactive rates, using only a realistic amount of collected data points.

Relative to analytical formulations for specific WPT configurations, our approach is complementary: closed-form models are extremely efficient within their assumptions (\textit{e.g.}, particular coil geometries and boundary conditions), whereas our model maintains interactive-speed prediction across diverse coil and shield configurations and under misalignment. The proposed framework, therefore, serves as a practical design-time tool when geometry or operating conditions depart from the regimes addressed by closed-form models.

In the preprocessing/parameterization stage, regarding geometry expression, our methods also excel at allowing regression with (i)~fixed-length vectors that facilitate GPR model construction, (ii)~compact vectors that simplify the model, and (iii)~linearizing the input-output relationship. In comparison, the multi-view projection~\cite{multiple-view} and voxel methods~\cite{voxel-based-modeling-1, voxel-based-modeling-2} usually support fixed-length vectors, while the signed distance field (SDF) on a Cartesian grid method~\cite{Modeling-SDF} (popular in fluid flow simulation) supports both fixed-length vectors and linear geometries. The point cloud approach~\cite{Modeling-pointnet} is known to support compact vectors primarily. Triangle mesh~\cite{Modeling-triangle-mesh} and boundary representation (B-Rep) methods~\cite{Modeling-b-req} both handle compact vectors and linear geometries. However, the length of these vectors changes as the geometry is modified, making it unsuitable for GPR (see Supplementary Table~2). For example, subdividing or reconnecting the triangle mesh is necessary for adding new edges and points.

Our adaptive exterior grid method also offers higher linearity between the inputs and outputs. Prior fixed grid-based methods~\cite{voxel-based-modeling-1, voxel-based-modeling-2, Modeling-voxel-cartesian} support fixed-length, compact geometry expression vectors but fall short in linearizing the input and output relationship. Meanwhile, the tetrahedra mesh methods~\cite{Modeling-tetra} support compact vectors and linear geometries, but the vector length varies with geometry modifications, limiting the application to GPR (see Supplementary Table~3). These comparisons highlight the comprehensive capability of our proposed method relative to existing approaches.

Regarding our GPR model with active learning, a comparison with other approaches is provided in Supplementary Table~1, showing a unique strength compared to conventional simulations or machine-learning approaches~\cite{Learning-WPT-gaussian}. Our proposed method offers significant improvements over prior methods in several key areas (see Supplementary Table~1 for details). While Knaish's method selects samples based on geometric proximity (\textit{e.g.}, linear distance in parameter space), the proposed active learning uses the model's predictive variance for data selection. It also expresses the interaction between the resonators and the field through adaptive 3D grids, leading to greater accuracy than prior work using fixed grids. These methods combined enable estimating the entire magnetic field accurately instead of a few points. Finally, our new method surpasses limitations in electromagnetic field solvers by enabling interactive estimations in \SI{130}{ms}.

\subsection*{Limitations and future work}
This study focused on \SI{85}{kHz}, a typical frequency for wireless charging of electric vehicles, and considered misalignments in the $yz$-plane within the range typically evaluated for such systems~\cite{sample2013enabling,fujita2017dynamic,xu2021dynamic} (see Methods). While this provides a realistic scope for automotive applications, extensions to MHz bands and broader 3D misalignments can be addressed by retraining the model with appropriately generated datasets, as the governing physical principles are the same; the effectiveness of electromagnetic field solvers across these frequency ranges suggests such extensions should be feasible. The framework also demonstrated robustness across varied coil and shield geometries (Supplementary Fig.~5), but substantially different coupler topologies will require retraining. In this study, training required about 700 full-wave simulations, which is modest compared with the $10^4$–$10^5$ solves needed for exhaustive sweeps.

We used full-wave electromagnetic simulations for validation, the practical standard in WPT design and widely relied upon to capture parasitic effects and near-field behavior. This provides an appropriate baseline for the scope of this study. An important next step is to incorporate experimental datasets and measurements.

Overall, these considerations represent practical boundaries rather than fundamental restrictions. The framework already enables interactive-speed field and efficiency estimation beyond the reach of conventional sweeps, and future work will extend the demonstrated scenarios to broader operating conditions and experimental settings.

%%%%%% Methods %%%%%%

\section*{Methods}
\subsection*{Data acquisition and simulation by FEKO}
To acquire the correct representation of magnetic field distribution and power transfer efficiency as training data for GPR, we use FEKO to perform calculations using the Method of Moments. The full-wave FEKO simulations used for training accounted for coil inductance, winding resistance, capacitance, electromagnetic losses, and parasitic effects (\textit{e.g.}, parasitic capacitance). Converter and peripheral circuit stages (\textit{e.g.}, rectifiers, DC–DC converters) were excluded, as they are typically modeled separately at the circuit level.

The coil shown in Supplementary Fig.~1a, b is used throughout the study. The inner and outer radii are \SI{90}{mm} and \SI{120}{mm}, respectively, with a structure of 28 turns in 4 layers (7 turns per layer). The height is set to \SI{15}{mm}. The material is copper, and the radius of the wire is \SI{1.6}{mm}. Additionally, magnetic resonance coupling is used as the wireless power transfer method, with a frequency of \SI{85}{kHz}. Capacitors were added to these coils so the transmitter and receiver resonated at this frequency. The input power is set individually for the shield's position and geometry based on the requirement for the receiver to receive \SI{1}{W} of power.
We defined power transfer efficiency as the maximum power transfer efficiency $\eta_\mathrm{max}$ obtained when the load value is adjusted to maximize power transfer efficiency~\cite{WPT-analysis-distance-etal}. This often leads to higher transmitter current and increased magnetic field exposure at lower power transfer efficiency, although the relationship is not strictly monotonic because the load value also changes~\cite{imura2020wireless}. The magnetic field distribution at the estimation points is calculated using trilinear interpolation from the values of the surrounding coordinates computed by MoM.

Furthermore, the shield geometry is shown in \fref{fig:complexity}a and Supplementary Table~5.
The magnetic shield is arranged to enclose the coil. The shield geometry is created using the parameters shown in \fref{fig:complexity}a.
The minimum values, maximum values, and intervals of these parameters are shown in Supplementary Table~5. However, depending on the relationship between $r_2$ and $d$, $r_1$ may exceed this limit, and such cases are excluded. Additionally, only $r_1$ and $h_1$ are parameterized when $h_2 = 0$. In this manner, 372 geometries are created. The relative position between the transmitter and receiver is shown in \fref{fig:complexity}b. The reference for the relative position between the transmitter and receiver is set to the position where there is no lateral misalignment, and the closest distance between the coils is \SI{5}{cm}. The relative position is defined as the movement from this reference point in the horizontal (y) direction and the vertical (z) direction. For the training data, the maximum values of y and z are set to \SI{24}{cm} (the coil diameter), and the intervals are set to \SI{2}{cm}. This value is empirically set, as previous studies on dynamic wireless power transfer typically evaluate performance up to about one coil diameter of misalignment~\cite{sample2013enabling,fujita2017dynamic,xu2021dynamic}. Larger misalignment leads to decreased efficiency and reduced practicality. In this way, a total of 169 relative positions are created. The circular shield is modeled with a triangular mesh, and the material is a ferromagnetic substance with a relative permeability of 3300 and a magnetic loss tangent of 0.001~\cite{WPT-shield-value}. 

\subsection*{Studied wireless power transfer system}

As shown in Supplementary Fig.~1a, the developed system employs a series–series resonant topology with a series resistive load at the receiver. For simplicity, this resistive load is commonly used to model downstream components such as rectifiers and DC–DC converters. The AC–to–AC (coil-to-coil) efficiency depends not only on the coil parameters and coupling coefficient but also on the load connected to the receiver. In wireless power studies, the load that maximizes efficiency is typically used, since it highlights the intrinsic performance of the coils themselves. Maximum efficiency point tracking techniques have been developed to maintain this condition~\cite{imura2020wireless}. The corresponding optimal efficiency $\eta_{\mathrm{opt}}$ is given in~\cite{Zargham2012, imura2020wireless} and is independent of the specific load value at the optimum. Accordingly, unless otherwise noted, reported efficiencies correspond to the optimal-load condition, $\eta_{\mathrm{opt}}$.

\subsection*{Construction of the aligned-edge polycube}
The construction of the aligned-edge polycube involves several key steps illustrated in \fref{fig:polycubemesh} to model complex geometries accurately:

\textit{Arrangement and Classification.}
We begin by arranging cubes into a polycube model that fits the target geometry, allowing us to classify vertices into corners (red dots) or surface points (pink squares) (\fref{fig:polycubeResult}). This classification facilitates subsequent placement and alignment adjustments.

\textit{Corner Point Adjustment.}
Corner points are carefully positioned to ensure alignment with neighboring plane structures. This involves adjusting their locations to maintain geometric integrity and alignment within the polycube framework (see Supplementary Fig.~6).

\textit{Surface Point Alignment.}
Surface points are adjusted based on corner point locations using interpolation techniques, ensuring a coherent alignment with the polycube's geometric planes (see also Supplementary Fig.~6). This step incorporates Mean Value Coordinates (MVC)~\cite{MVC} to fine-tune the alignment with the object's surface.

\textit{Subdivision for Detail.}
We subdivide the initial grid, introducing additional points at strategic locations such as edge midpoints and face centroids (\fref{fig:polycubemesh}a).

\textit{Edge Projection.}
Finally, edge projection ensures precise alignment by iteratively refining point positions. Projections are carried out perpendicularly and parallelly to existing structures to achieve an optimal fit, even in spaces constrained by geometric complexity (Supplementary Fig.~7).

\subsection*{Construction of the adaptive exterior grid}
A grid is set (see in blue in \fref{fig:polycubemesh}b) around the shield-fitting-cubes (polycubes in green on the same figure), with intersections corresponding to our sampling points. We define the points belonging to the grid as `exterior points'. Next, we project those exterior points onto the shield geometry. We use the 3D extension of the Linear Blend Skinning (LBS) method~\cite{LBS-linear-blend-skinning} to adjust their positions linearly based on the locations of all the corner points. The LBS calculates the weights that translate the movement of a corner point to the movement of our exterior points (the `adaptive' part of our method). From there, we subdivide this exterior cubic grid. While it would be possible to selectively subdivide regions of the grid, in this study, we divide the grid uniformly. The new points are placed at the centroids of each grid cube face and edge midsections.

The LBS method works by blending multiple transforms and is generally used in computer graphics for fast rendering. First, let the coordinates of $n$ points on two shields be $p_1,\cdots, p_n$. Also, if the coordinates of one estimation point are $p_e$, the weights of each point on the ferrite shield for this estimation point can be defined as follows.
\begin{linenomath}
\begin{align}
    w_i = \dfrac{\dfrac{1}{|p_i - p_e|^d}}{\displaystyle \sum_{k = 1}^{n} \dfrac{1}{|p_k - p_e|^d}}
\end{align}
\end{linenomath}
Here, $d$ is calculated as $d = 1.5$ for geometry changes of the ferrite shield and $d = 5.0$ for shield movements. This weight represents how close each point on the ferrite shield is to that estimation point (the value rises as it gets closer). By summing the change amounts $q_i$ of the points on the shield's surface (for this specific weight), the change amount $q_e$ of the estimation point can be calculated. Otherwise put, it becomes as follows:
\begin{linenomath}
\begin{align}
    q_e = \displaystyle \sum_{i = 1}^{n} w_i * q_i
\end{align}
\end{linenomath}

By performing this calculation for each point, we can determine the coordinates of the destination point. In this study, we use 72 points at subdivision level 1 (initial shield-fitting-polycubes) for each shield (144 points in total for both shields).

\subsection*{Selecting estimation points for active learning}
Here, we detail our strategy to increase the amount of training data, as illustrated in \fref{fig:activelearning}. We develop a Gaussian Process Regression (GPR) model using an initial training dataset of 744 data points, producing model instances corresponding to each estimation point. Subsequently, these model instances are employed to calculate the field for all data points within the unlabeled dataset.

Our approach identifies six model instances based on the highest average deviation observed in these calculations. To avoid selecting data points that are spatially too close, we clustered them in a 3$\times$3$\times$3 grid (27 regions, see Supplementary Fig.~8a, b). This process allows us to identify six challenging estimation points susceptible to errors.

The dataset generation is then done iteratively until the training dataset size is doubled to 1,488 data points. The following steps summarize our strategy: (i) the remaining unlabeled data points are processed through the six selected model instances, (ii) for each model instance, the data point with the highest deviation is identified, resulting in the selection of six data points, and (iii) these selected data points are then simulated using FEKO and integrated into the training dataset.

\subsection*{Gaussian process regression}
We use GPR as a regression method. GPR generates estimations based on the the weighted sum of the distances between the input (here, shields geometries and their relative positions) and the target data (corresponding magnetic field and power transfer efficiency). Our motivation for using the GPR method in this study is threefold: (i) the method can estimate the nonlinear relationship between inputs and outputs with a small number of training data, (ii) the GPR method provides us with a confidence metrics  as well, so we can gauge the estimation results, and (iii) this method therefore enables us to identify which samples in the training data can be used reliably. GPR is helpful in cases where creating a large dataset would be unpractical, such as with WPT simulations: it has been shown that GPR can achieve high-accuracy estimations even with limited training data~\cite{GPR-2, GPR-3}. Additionally, since GPR is a probabilistic model, the method allows us to compute the variance of the estimated distribution, so that we can further assess the confidence in the estimations. With this confidence measure, we can then add samples to low-confidence areas utilized in the active learning process described in its dedicated section.

GPR is a regression method that assumes both the input $\mathbf{x}$ and the output $\mathbf{y}$ follow a Gaussian distribution, and it uses a kernel function $k_{GPR}$ to make more flexible estimations.

An overview explanation of the method has been provided~\cite{GPR-1, GPR-explanation}; here, we will detail the mathematical formulation of GPR~\cite{Learning-fluid-umetani-2}.
Suppose we have training data $\mathcal{D} = \{(\mathbf{x}_n, y_n)\}^N_{n = 1}$ for the output $\mathbf{y}$ given the input $\mathbf{x}$.
However, it is desirable to have the expected value of the output equals to zero $\mathbb{E}[y] = 0$.
Here, we denote the concatenation of the outputs in the training data $y_n$ as $\mathbf{y} = \{y_1,\cdots, y_n\}$. The GPR assumes that this output $\mathbf{y}$ follows a multivariate Gaussian distribution $\mathbf{y} \sim \mathcal{N}(\mathbf{0},\mathbf{K_{GPR}})$, where $\mathbf{K_{GPR}}$ is the covariance matrix of $\mathbf{y}$.
Here, using the output vector $\mathbf{y}$, the covariance matrix $\mathbf{K_{GPR}}$ is defined as $\mathbf{K_{GPR}} = \mathbb{E}[\mathbf{y}\mathbf{y}^T]$. Each element of $K_{GPRij}$ indicates the magnitude of the correlation between the output $y_i$ and $y_j$, and in GPR, the covariance matrix is modeled using the kernel function $K_\mathrm{GPRij} = k_\mathrm{GPR}(\mathbf{x}_i, \mathbf{x}_j)$.

Next, we consider the output $y^*$ for a new input $\mathbf{x}^*$.
We denote the concatenation of the outputs $\mathbf{y}$ and the unknown output $y^*$ as $\mathbf{y}^\prime = \{y_1,\cdots, y_n, y^*\}$. 
If we define $\mathbf{K_{GPR}}^\prime$ as the $(N+1)\times(N+1)$-dimensional covariance matrix calculated from $\mathbf{x}_1, \cdots,\mathbf{x}_N$ and $\mathbf{x}^*$, this also follows a Gaussian distribution, so it becomes $\mathbf{y}^\prime \sim \mathcal{N}(\mathbf{0},\mathbf{K_{GPR}}^\prime)$. Otherwise put:
\begin{linenomath}
\begin{align}
    \begin{pmatrix}
        \mathbf{y} \\
        y^* \\
    \end{pmatrix}
    \sim \mathcal{N}\left(\mathbf{0},
    \begin{pmatrix}
        \mathbf{K_{GPR}} & \mathbf{k_{GPR}}_* \\
        \mathbf{k_{GPR}}_*^T & k_\mathrm{GPR**} \\
    \end{pmatrix}\right)~.
\end{align}
\end{linenomath}
In this formula, $\mathbf{k_{GPR}}_* = \left(k_\mathrm{GPR}(\mathbf{x}^*, \mathbf{x}_1), \cdots, k_\mathrm{GPR}(\mathbf{x}^*, \mathbf{x}_N)\right)^T$ is the vectors that list the similarity between the new input $\mathbf{x}^*$ and the input of the training data, and $k_\mathrm{GPR**} = k_\mathrm{GPR}(\mathbf{x}^*, \mathbf{x}^*)$ is the similarity between $\mathbf{x}^*$ and itself.
Using these, the probability distribution of the unknown output $y^*$ (which is an element of vector $\mathbf{y^*}$) can be expressed as follows
\begin{linenomath}
\begin{align}
    p(y^*|\mathbf{x}^*,\mathcal{D}) &= \mathcal{N}(\mathbf{w}^*\mathbf{y}, k_\mathrm{GPR**}-\mathbf{w}^*\mathbf{k_{GPR}}_*)\\
    \mathbf{w}^* &= \mathbf{k_{GPR}}_*^T\mathbf{K_{GPR}}^{-1}
\end{align}
\end{linenomath}
We apply GPR to the regression of power transfer efficiency.
In GPR, the expected value of the output is required to be zero, so we subtract the average $\tilde{\eta}$ from the training data for power transfer efficiency and offset it to obtain $y = \eta - 	\tilde{\eta}$.
\begin{linenomath}
\begin{align}
    \eta(\mathbf{x}^*) = \mathbf{w}^* \boldsymbol{\eta} + w^* \tilde{\eta},~~\text{where}~~w^* = 1-\sum_{n = 1}^{N}w_n^*
\end{align}
\end{linenomath}
Here, $\boldsymbol{\eta}$ is a vector that concatenates the $\eta$ values of the training data.

In the same way, we also perform regression on the absolute value $\mathbf{h}$ and the vector direction $\mathbf{v}_\theta$, $\mathbf{v}_\varphi$ of the magnetic field distribution. We define the vector using the zenith angle $\mathbf{v}_\theta$ from the z-axis and the azimuth angle $\mathbf{v}_\varphi$ from the x-axis in the spherical coordinate system.
\begin{linenomath}
\begin{align}
    \mathbf{h}(\mathbf{x}^*) &= \mathbf{w}^* \mathbf{H} + w^* \tilde{\mathbf{h}} \\
    \mathbf{v}_\theta(\mathbf{x}^*) &= \mathbf{w}^* \mathbf{V}_\theta + w^* \tilde{\mathbf{v}}_\theta \\
    \mathbf{v}_\varphi(\mathbf{x}^*) &= \mathbf{w}^* \mathbf{V}_\varphi + w^* \tilde{\mathbf{v}}_\varphi
\end{align}
\end{linenomath}
Here $\tilde{\mathbf{h}}$, $\tilde{\mathbf{v}}_\theta$, and $\tilde{\mathbf{v}}_\varphi$  are column vectors representing the row-wise averages of $\mathbf{H}$, $\mathbf{V}_\theta$, and $\mathbf{V}_\varphi$, respectively, which are matrices stacking training data in rows.

The weight $\mathbf{w}^*$ is identical to the one used for power transfer efficiency. Ideally, a distinct weight would be prepared for each estimation point, but this would lead to an excessively large estimation model, making the learning process time-consuming beyond practical limits. Moreover, having numerous models would increase the predicted time due to computer memory constraints. Additionally, since power transfer efficiency has a substantial impact on magnetic field distribution, it is logical to perform these calculations using the weight designated for power transfer efficiency. We also observe that, based on a similar rationale, previous research in fluid dynamics utilized weights computed from the drag coefficient to accurately estimate velocity and pressure at each point~\cite{Learning-fluid-umetani-2}.

Finally, we selected the following kernel functions.
\begin{linenomath}
\begin{align}
    k_\mathrm{GPR}(\mathbf{x},\mathbf{x}^{\prime}) = \theta_a\exp\biggl{(}-{\displaystyle \sum_{i = 1}^{M}}\frac{(\mathbf{x}_i - \mathbf{x}_i^{\prime})^2}{\theta_i}\biggr{)} + \theta_b\delta(\mathbf{x},\mathbf{x}^{\prime})
    \label{eq:kernel}
\end{align}
\end{linenomath}
The first term in equation (\ref{eq:kernel}) defines the similarity between $\mathbf{x}$ and $\mathbf{x}^\prime$ in terms of the similarity between the outputs $\mathbf{y}$ and $\mathbf{y}^\prime$. The hyperparameter $\theta_a > 0$ is a positive uniform scaling factor, and $\boldsymbol{\theta} = (\theta_1, \cdots, \theta_M) > 0$ is a positive scaling factor for each dimension of the input.
We choose this squared exponential covariance function (kernel function) because it corresponds to a linear regression with an infinite number of basis functions.
The second term in equation (\ref{eq:kernel}) models the Gaussian white noise with variance $\theta_b > 0$ assumed in the output.
Here, we optimize these hyperparameters $\boldsymbol{\Theta} = \{\theta_a, \boldsymbol{\theta}, \theta_b\}$ using maximum likelihood method on power transfer efficiency.
\begin{linenomath}
\begin{gather}
    \mathrm{arg}~\underset{\boldsymbol{\Theta}}{\mathrm{max}}~p(\mathbf{y}|\boldsymbol{\Theta}) = -\mathcal{L}(\boldsymbol{\Theta}) \\
    \mathcal{L}(\boldsymbol{\Theta}) 
    = \frac{N}{2} \log|\mathbf{K_{GPR}(\boldsymbol{\Theta})}| + \frac{1}{2}\mathbf{y}^T\mathbf{K_{GPR}}^{-1}(\boldsymbol{\Theta})\mathbf{y} + \frac{N}{2}\log(2\pi)
\end{gather}
\end{linenomath}
This $\mathcal{L}(\boldsymbol{\Theta})$ is minimized using the gradient descent method over 1,500 epochs.

\section*{Data availability}
The dataset supporting this study is archived on Zenodo (Dataset v0.1.0; DOI: \href{https://doi.org/10.5281/zenodo.17397941}{10.5281/zenodo.17397941})~\cite{honjo_2025_dataset}.

\section*{Code availability}
The code used in this study is archived on Zenodo (Software v0.1.3; DOI: \href{https://doi.org/10.5281/zenodo.17378264}{10.5281/zenodo.17378264})~\cite{honjo_2025_code}, and mirrored at \url{https://github.com/SasataniLab/EMFieldGPR}.

%%%%%% Reference %%%%%%
\printbibliography

%%%%%% Other information %%%%%%
\section*{Acknowledgements}
This work was supported by JST FOREST Program (Grant Number JPMJFR242P) and by JSPS KAKENHI (Grant Numbers 23K28068 and 23H03378).

\section*{Author contributions}
Y.H. and T.S. designed the research.
Y.H. conceived the idea of this study, developed the modeling methods, and analyzed the results.
Y.H. and T.S. developed the software.
K.T. and C.C. provided the ideas necessary for model creation. 
Y.N. provided the ideas necessary for modeling.
Y.H., T.S., and C.C. wrote the manuscript.
K.T. revised the manuscript.
Y.H., C.C., K.T., Y.N., Y.K., and T.S. reviewed and approved the final manuscript.
T.S. acquired funding and supervised the project.

\section*{Competing interests}
The authors declare no competing interests.

%%%%%% Figures %%%%%%
\newgeometry{left=0.66in, right=0.66in, top=0.66in, bottom=0.66in}
\newpage

\newpage

\newpage

\newpage
\begin{figure}[h]
    \centering
    \includegraphics[width=84mm]{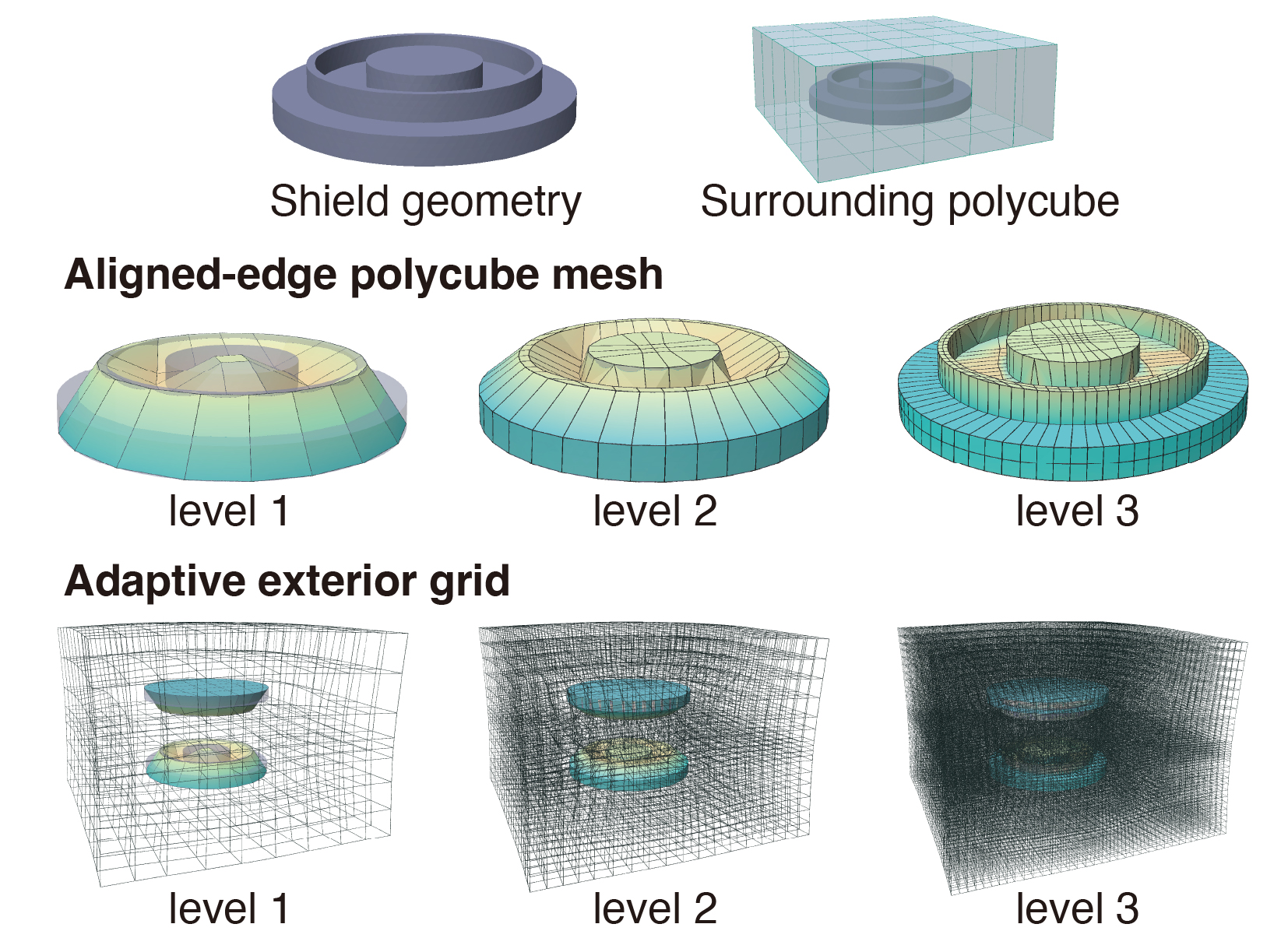}
    \caption{Generated aligned-edge polycube and adaptive exterior grid. The aligned-edge polycubes are specifically aligned with the corners of the shield geometry, while the adaptive exterior grid points are created to interface with the polycube meshes. After analyzing the resulting mesh, it was concluded that a level 3 aligned-edge polycube sufficiently represents the geometry, and a level 2 adaptive exterior grid effectively captures the magnetic field. Additional evaluations at level 3 were conducted to determine the effect on processing time by examining more detailed aspects of the field distribution.}
    \label{fig:polycubeResult}
\end{figure}

\newpage
\begin{figure}[h]
    \centering
    \includegraphics[width=176.5mm]{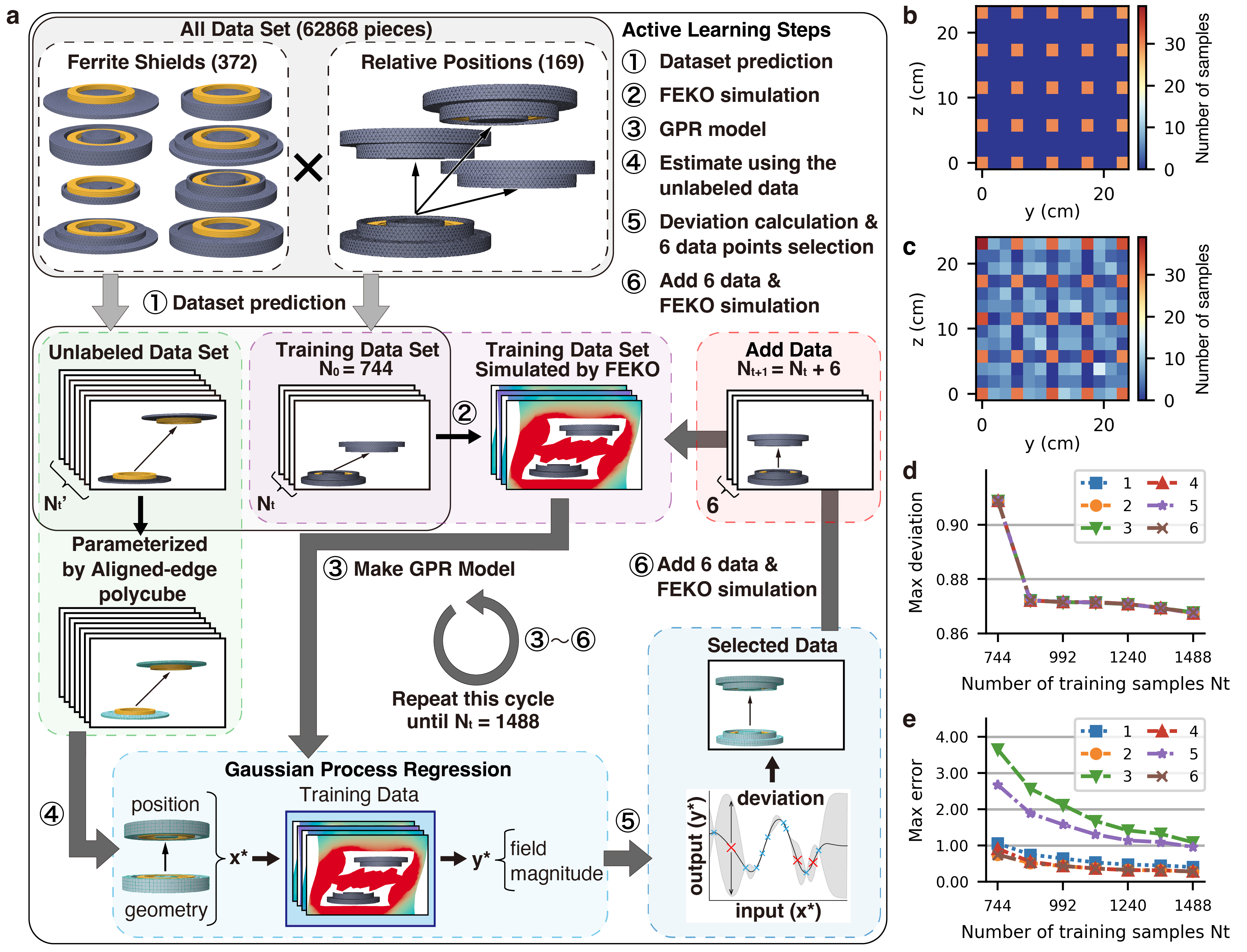}
    \caption{Training dataset construction using active learning for GPR.
    \textbf{a} Overview of the active learning process using GPR. Initially, we created an unlabeled dataset consisting of 169 relative positions for each of the 372 shields. From this dataset, 744 data points were simulated to form the initial training dataset, while the remaining data points were left unlabeled. The entire unlabeled dataset was then analyzed to identify six data points corresponding to GPR model instances showing the highest average deviation, and these were used throughout the active learning process. For each cycle, we calculated the deviation of each data point using these GPR instances, selected those with significant deviations to add to the training set, and repeated the cycle until the training dataset size was doubled to 1,488 data points.
    \textbf{b} Distribution of the relative positions in the initial training dataset ($N_0=744$).
    \textbf{c} Distribution of the relative positions in the training dataset after active learning ($N_\mathrm{t}=1488$).
    \textbf{d,e} Relationship between training dataset size, maximum deviation, and maximum error of selected GPR model instances at the start of active learning. Both maximum deviation and maximum error decrease as the training dataset size increases.}
    \label{fig:activelearning}
\end{figure}

\newpage
\begin{figure}[h]
    \centering
    \includegraphics[width=84mm]{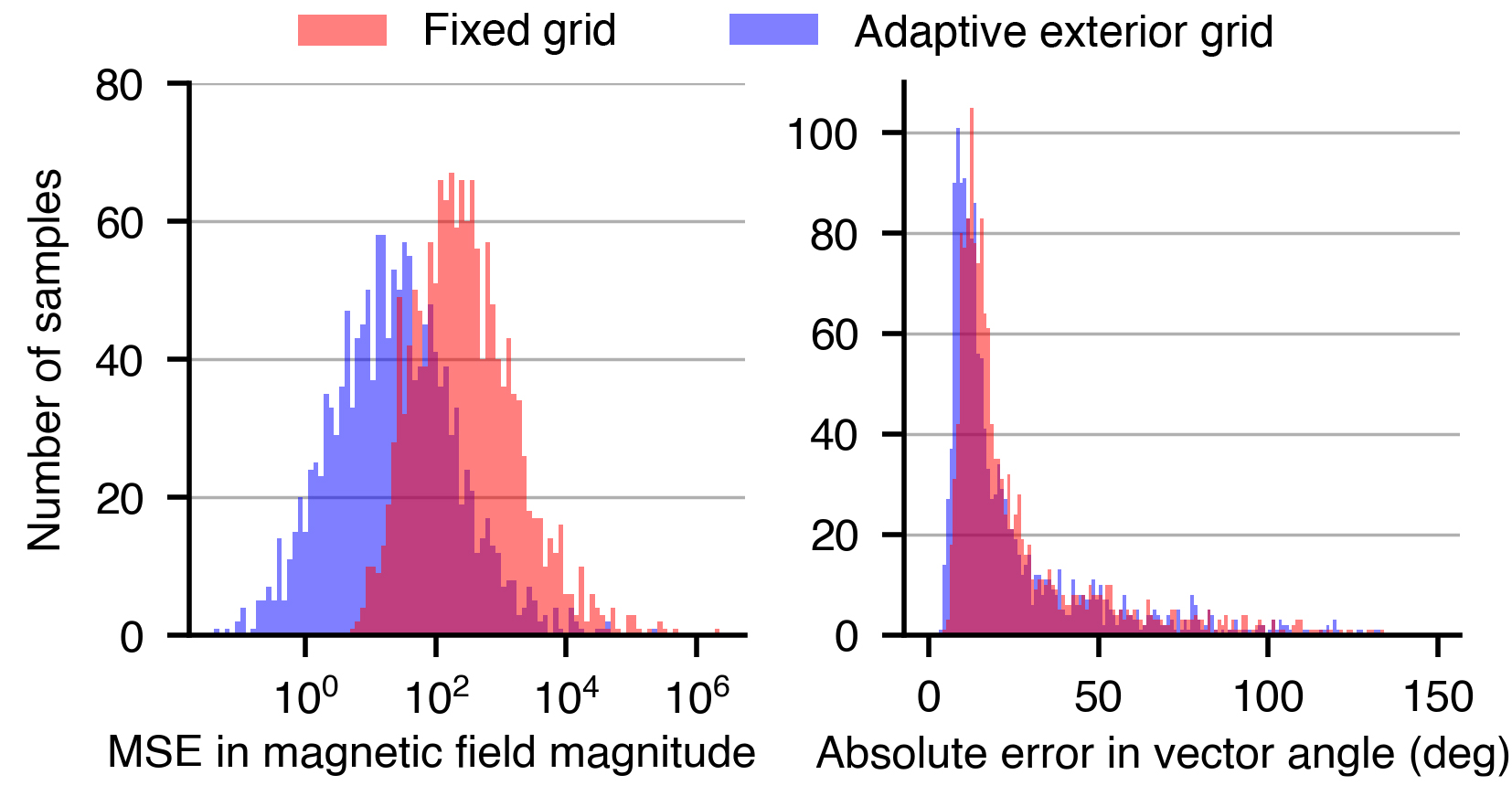}
    \caption{Estimation error with the proposed adaptive exterior grid and fixed grid. The magnetic field magnitude is evaluated using mean squared error, while the vector estimation is assessed based on the error in absolute length. Both metrics show improvement with the proposed adaptive grid, with a particularly notable enhancement in the accuracy of the magnitude estimation. In this evaluation, cross-validation is applied.}
    \label{fig:evaluationExteriorGrid}
\end{figure}

\clearpage
\begin{figure}[h]
    \centering
    \includegraphics[width=176.5mm]{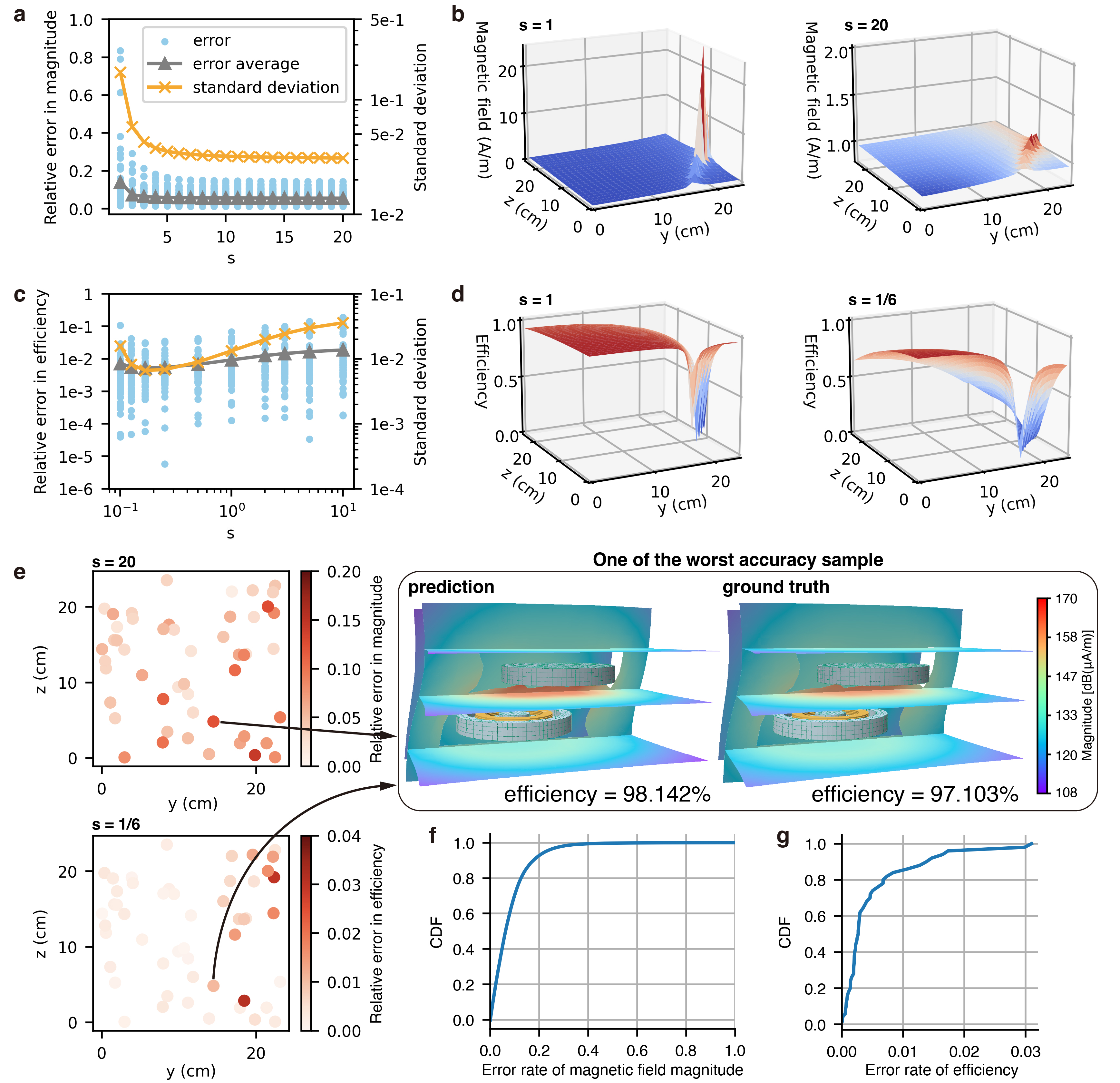}
    \caption{Evaluation of the postprocessing method for improving linearity.
    \textbf{a} Magnetic field magnitude estimation error when raising the data points to the power of $s$. As $s$ increases, both the average estimation error and deviation decrease, eventually converging when $s$ becomes sufficiently large.
    \textbf{b} Visualization of changes in the magnetic field due to the coil's relative position at $s=1$ and $s=20$, indicating that postprocessing mitigates steep changes in field magnitude with positional shifts.
    \textbf{c} Power transfer efficiency estimation error when raising the data points to the power of $1/s$. The minimum average and standard deviation of these errors occur at $s=6$.
    \textbf{d} Comparison of power transfer efficiency changes due to coil position at $s=1$ and $s = 1/6$, illustrating how this transformation mitigates steep efficiency variations.
    \textbf{e} Error map for magnetic field and power transfer efficiency estimation with best-performing postprocessing parameters, alongside a visualization of a low-accuracy magnetic field estimation case.
    \textbf{f} Cumulative distribution function (CDF) of the relative error from ground truth data for magnetic field magnitude estimation. 80\% of data points have an error of up to 12\%.
    \textbf{g} CDF of the relative error from ground truth data for power transfer efficiency estimation, with 80\% of data points having an error of up to 0.7\%.}
    \label{fig:pow}
\end{figure}

\clearpage
\begin{figure}[h]
    \centering
    \includegraphics[width=176.5mm]{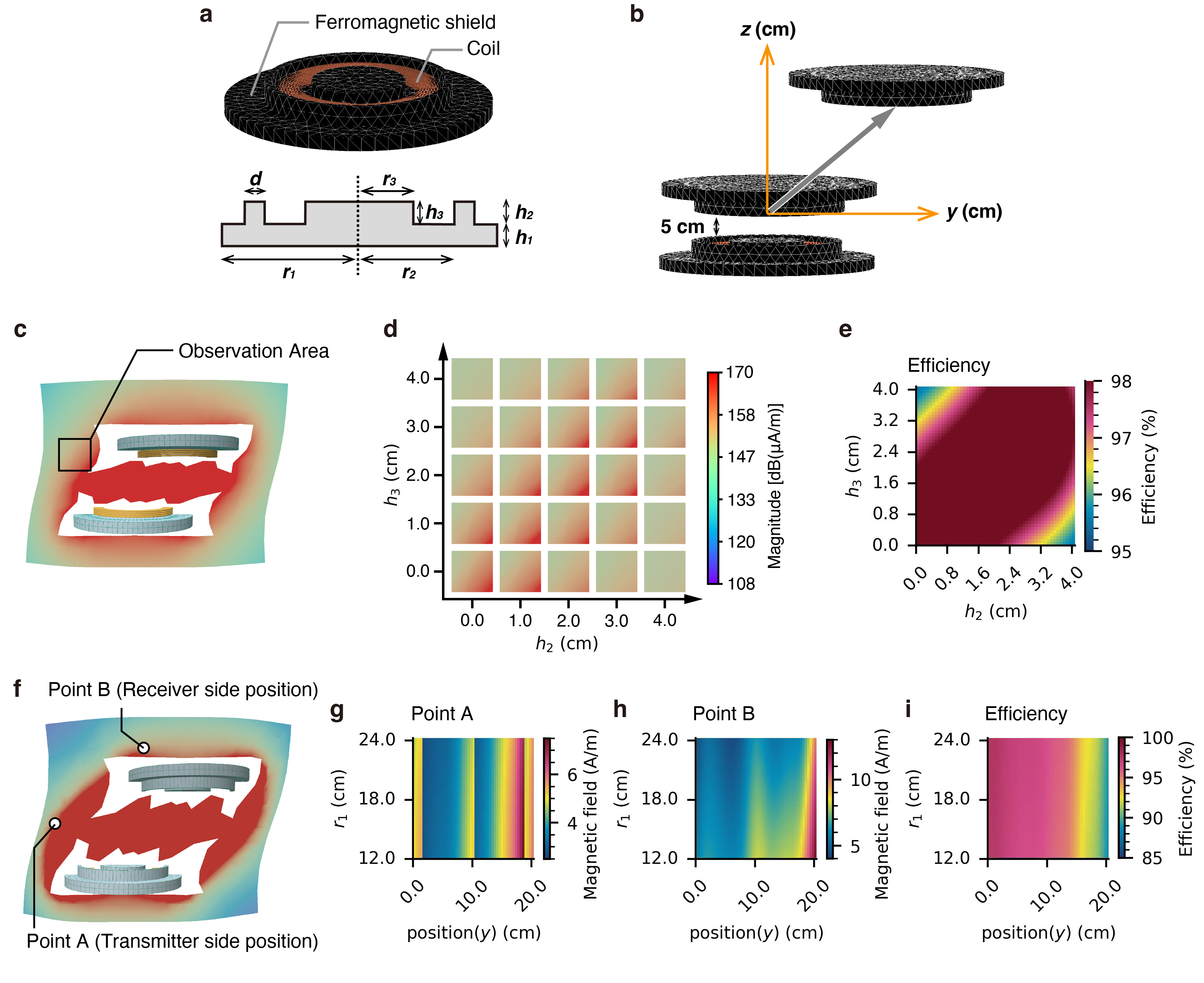}
    \caption{An illustrative example of interactive parameter exploration using arrays of outputs. 
    These outputs provide rapid insight into coupled, nonlinear dependencies among geometry, position, and performance, showing how shield geometry and misalignment influence magnetic-field distribution and coil-to-coil efficiency. 
    \textbf{a} Geometry and design parameters ($r_1, r_2, r_3, h_1, h_2, h_3, d$). 
    \textbf{b} Variations in the relative position between the transmitter and receiver, 
    with the reference defined as having no lateral misalignment and a separation of \SI{5}{cm}. 
    Positional deviations are expressed as displacements in the horizontal ($y$) and vertical ($z$) directions from this reference position. 
    \textbf{c–e} Example of a static wireless power case. 
    \textbf{c} Observation area for the magnetic field, where shield heights $h_2$ and $h_3$ are swept. 
    The receiver coil position is fixed at $y=\SI{13.0}{cm}$, $z=\SI{11.0}{cm}$. 
    \textbf{d} Magnetic field magnitude within the observation region and 
    \textbf{e} corresponding efficiency in the static case. 
    \textbf{f–i} Example of a dynamic wireless power case. 
    \textbf{f} Definition of evaluation points, where Point~A is fixed at the transmitter side and Point~B moves with the receiver. 
    In this case, the coil lateral misalignment $y$ and inner-shield radius $r_1$ are swept. 
    The coil’s vertical position is fixed at \SI{20}{cm}; Point~A is at $y=\SI{-24.5}{cm}$, $z=\SI{12}{cm}$, and Point~B at $y=\SI{-15.0}{cm}$, $z=\SI{36}{cm}$. 
    \textbf{g} Magnetic-field magnitude at Point~A, \textbf{h} at Point~B, and \textbf{i} efficiency in the dynamic case.}

\label{fig:complexity}
\end{figure}

\clearpage
\begin{table}[t]
    \caption{Detail level of the exterior grid and its processing speed.}
    \label{table:speed}
    \centering
    \begin{tabular}{wc{28mm} | wr{18mm} | wr{15mm} | wr{15mm}}
        \hline
        Adaptive Exterior &  \multicolumn{1}{c|}{ Number of}   & \multicolumn{2}{c}{Processing Speed} \\
        \cline{3-4}
        Grid Level & \multicolumn{1}{c|}{points} & \multicolumn{1}{c|}{position} & \multicolumn{1}{c}{geometry} \\
        \hline
        \rowcolor{gray!0}
        Level 1              & 756   & \SI{31}{ms}   & \SI{37}{ms}  \\
        Level 2              & 5411    & \SI{96}{ms}   & \SI{130}{ms}  \\
        Level 3              & 40767    & \SI{688}{ms}  & \SI{1287}{ms} \\
        \hline
    \end{tabular}
\end{table}

\end{document}